\documentclass[3p,times,procedia]{elsarticle}

\usepackage{ecrc}
\usepackage{algorithm}
\usepackage{algorithmic}
\usepackage{subfigure}

\volume{00}

\firstpage{1}

\journalname{Procedia Computer Science}

\runauth{}

\jid{procs}

\jnltitlelogo{Procedia Computer Science}

\CopyrightLine{2011}{Published by Elsevier Ltd.}

\usepackage{amssymb}

\usepackage[figuresright]{rotating}

\begin{document}

\begin{frontmatter}

\dochead{}

\title{Q-LEACH: A New Routing Protocol for WSNs}

\author{B. Manzoor$^{\pounds}$, N. Javaid$^{\pounds}$, O. Rehman$^{\pounds}$, M. Akbar$^{\pounds}$, Q. Nadeem$^{\pounds}$, A. Iqbal$^{\pounds}$, M. Ishfaq$^{\S}$}

\address{$^{\pounds}$COMSATS Institute of Information Technology, Islamabad, Pakistan. \\
        $^{\S}$King Abdulaziz University, Rabigh, Saudi Arabia.}

\address{}

\begin{abstract}
Wireless Sensor Networks (WSNs) with their dynamic applications gained a tremendous attention of researchers. Constant monitoring of critical situations attracted researchers to utilize WSNs at vast platforms. The main focus in WSNs is to enhance network life-time as much as one could, for efficient and optimal utilization of resources. Different approaches based upon clustering are proposed for optimum functionality. Network life-time is always related with energy of sensor nodes deployed at remote areas for constant and fault tolerant monitoring. In this work, we propose Quadrature-LEACH (Q-LEACH) for homogenous networks which enhances stability period, network life-time and throughput quiet significantly.
\end{abstract}


\end{frontmatter}


\label{}
\section{Background and Motivation}
WSNs are considered one of the best sources for monitoring remote fields and critical conditions which are out of range from humans perspective. For optimal distribution of energy among sensor nodes, in order to enhance network life time, suitable protocols and applications should be developed.

Based upon optimal probability, selection of cluster heads (CHs) is discussed in homogenous clustering protocol called Low Energy Adaptive Cluster Hierarchy (LEACH) [1], for load distribution of energy within sensors. Moreover, concept of hierarchal and multi-hop clustering distributes energy load more evenly. It is noticed that localized schemes perform well when compared with centralized algorithm in clustering based approaches.

On the basis of energy distribution among sensor nodes, WSNs are classified into homogenous and heterogenous networks. Some clustering protocols such as LEACH [1], Power-Efficient Gathering in Sensor Information System (PEGASIS) [2], and Hybrid Energy-Efficient Distributed Clustering (HEED) [3] are defined for homogenous networks. Whereas, stable Election Protocol (SEP) [4] and Distributed Energy-Efficient Clustering (DEEC) [5] deal with heterogeneous networks.

Through geographical information and energy awareness of nodes, Geographic and Energy Aware Routing (GEAR) [6] routes a packet towards targeted region. For such process either their exist a closer neighbor or all neighbor are farther away from destination. For closer neighbors from the destination, GEAR picks a next-hope node among all neighbors closer to the destination. In case of distant neighbors their exists a hole and GEAR selects a next-hope node on the basis of minimum cost value. Moreover, Energy Aware Geographic Routing Protocol (EAGRP) is another technique used in wireless networks for routing packets [7].

Sensor networks are deployed for long term monitoring of fields and are desired to continue working without abrupt changes. Moreover, it is also
desired to obtain global knowledge continuously i.e., better coverage of area should be obtained. Considering above mentioned needs new approach Q-LEACH
is designed which improves network efficiency. Remainder of this paper is as follows: Section II describes our proposed model for efficient energy utilization in WSNs. Simulation results are discussed in section III, and finally section IV concludes the paper.

\section{Q-LEACH}
In this section, we discuss our proposed strategy named as Q-LEACH. We discuss network characteristics and working principle of proposed scheme for efficient performance. This section presents key concept of proposed network model. In order to enhance some features like clustering process, stability period and network life-time for optimized performance of WSNs we propose this model.

\textbf{A}ccording to this approach sensor nodes are deployed in the territory. In order to acquire better clustering we partition the network into four quadrants. Doing such sort of partitioning better coverage of the whole network is achieved. Additionally, exact distribution of nodes in field is also well defined.

\begin{figure}
  \begin{centering}
\includegraphics[height=5 cm,width=5 cm]{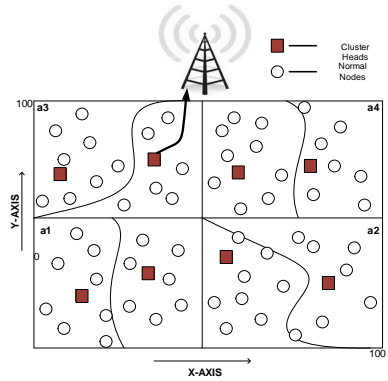}
 \caption{Network Topology}
 \end{centering}
\end{figure}

Fig.1 describes optimal approach of load distribution among sensor nodes. Moreover, it also presents an idea of efficient clustering mechanism which yields significantly in better coverage of whole network. We deployed random nodes in a $100m\times100$m filed. Based on location information, network is divided into four equal parts i.e, $(a1,a2,a3,a4)$. Defining overall network area as below:

\begin{eqnarray}
A=a_1+a_2+a_3+a_4
 \end{eqnarray}

 $a_n=A(x_m,y_m)$

 Where, $n=4$. and $m=100$.
 Hence, overall field is distributed as follows:
 \begin{eqnarray}
\lim_{X_m=0:50}^{Y_m=0:50} a_n + \lim_{X_m=51:100}^{Y_m=0:50} a_n + \lim_{X_m=0:50}^{Y_m=51:100} a_n +\lim_{X_m=51:100}^{Y_m=51:100} a_n
 \end{eqnarray}
Portioning of network into quadrants yields in efficient energy utilization of sensor nodes. Through this division optimum positions of CHs are defined. Moreover, transmission load of other sending nodes is also reduced. In conventional LEACH cluster are arbitrary in size and some of the cluster members are located far away. Due to this dynamic cluster formation farther nodes suffers through high energy drainage and thus, network performance degrades. Whereas, in Q-LEACH network is partitioned into sub-sectors and hence, clusters formed within these sub-sectors are more deterministic in nature. Therefore, nodes are well distributed within a specific cluster and results in efficient energy drainage. Concept of randomized clustering as given in [1] for optimized energy drainage is applied in each sector. Assigning CH probability $P=0.05$ we start clustering process. In every individual round nodes decides to become CH based upon \textit{P} and threshold \textit{T(n)} given in [1] as:
\begin{algorithm}[H]
\caption{Setup Phase}
\begin{algorithmic}[1]
\STATE{begin}
\IF{node $ \varepsilon G\longrightarrow G= $nodes which did not become CHs in current EPOCH}
       \IF{$( NODE\_BELONGS\_TO == 'areaA')$ }
       \IF{$( NUMBER OF CHs <= \left(\frac{N}{K} \right))$ }
       \STATE       TEMP=random number (0-1)
       \IF{$(temp <= \frac{P}{1-P(r,mod1/P)})$}
       \STATE node=CH\_A
       \STATE NUMBER\_OF\_CHs =  NUMBER\_OF\_CHs+1
       \ENDIF
       \ELSIF{$( NODE\_BELONGS\_TO == 'areaB')$ }
       \STATE    REPEAT STEP 4 : 8
       \ELSIF{$( NODE\_BELONGS\_TO == 'areaC')$ }
       \STATE    REPEAT STEP 4 : 8
       \ELSIF{$( NODE\_BELONGS\_TO == 'areaD')$ }
       \STATE    REPEAT STEP 4 : 8
       \ENDIF
       \ENDIF
       \ENDIF
 \end{algorithmic}
 \end{algorithm}
 Algorithm.1 defines CHs selection mechanism. Overall network is divided into four areas as: Area A, B, C and D. Initially each node decides whether or not to become a CH. Node chooses a random number between $0$ and $1$. If this number is less then certain threshold \textit{T(n)}, and condition for desired number of CHs in a specific area is not met, then the node becomes a CH. Similarly the same process continues for rest of the sectors and optimum number of clusters are formed. Selection of clusters will depend upon Received Signal Strength Indicator (RSSI) of advertisement. After decision of clusters, nodes must tell CHs about their association. On the basis of gathered information from attached nodes, guaranteed time slots are allocated to nodes using Time Division Multiple Access (TDMA) approach. Moreover this information is again broadcasted to sensor nodes in the cluster.

 Algorithm.2 defines association of nodes with their appropriate CHs. Non-CHs nodes will locate themselves in specified area they belong to. Then they will search for all possible CHs, and on the basis of RSSI they will start association. This process will continue until association phase comes to an end.

Once cluster setup phase is complete and nodes are assigned with TDMA slots every node communicates at its allocated time interval. Rest of the time radio of each non-cluster head node will remain off in order to optimize energy utilization. When all nodes data is received at the CHs then, the data is compressed and is sent to BS. The round completes and new selection of CHs will be initiated for next round.

In proposed idea, we implement above mentioned concept of localized coordination in each sectored area. We used same radio model as discussed in [1] for transmission and reception of information from sensor nodes to $CHs$ and then to $BS$. Packet length $K$ of $2000$ bits is used in our simulations.

According to above mentioned flow chart, initially all nodes send their location information to BS. BS performs logical partitioning of network on the basis of gathered information. Network is divided into four quadrants and broadcasts information to nodes. On the basis of threshold some nodes are elected as CH in each division. Normal nodes choose their CHs within their own quadrant based on RSSI. For association nodes sends their requests to CHs. TDMA slots are assigned to every node for appropriate communication without congestion. Every node communicates in its allocated slot with its defined CH.
\begin{algorithm}[H]
\caption{Node Association in Q-LEACH}
\begin{algorithmic}[1]
    \STATE $N \in $ $Group$ $of$ $normal$ $nodes$
     \STATE $GC \in $ $Group$ $of$ $CHs$
    \IF{$N$ $\in$ $(A,a_1)$}
    \STATE $Where$
    \STATE $A$ $=$ $a_1$,$a_2$,$a_3$,$a_4$
    \STATE Check all possible ACHs
    \STATE Check RSSI of CHs
    \STATE Associate with ACHs
    \STATE then
    \STATE transfer of data occurs
     \ENDIF
     \IF{$N$ $\in$ $(A,a_2)$}
     \STATE Repeat step from 5 : 8 for BCHs
\ENDIF
     \IF{$N$ $\in$ $(A,a_3)$}
     \STATE Repeat step from 5 : 8 for CCHs
\ENDIF
     \IF{$N$ $\in$ $(A,a_4)$}
     \STATE Repeat step from 5 : 8 for DCHs
\ENDIF
     \end{algorithmic}
\end{algorithm}

\begin{figure}
\centering
{\includegraphics[height=8 cm,width=6 cm]{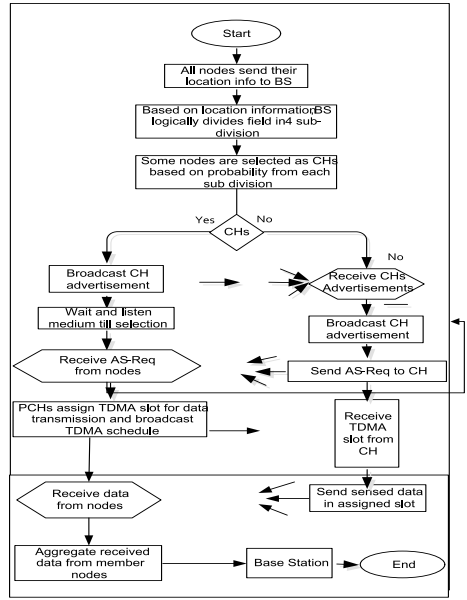}}
\caption{Working Principle of Q-LEACH }
\end{figure}

\section{Simulation Results}
In this section, we discuss and compare simulations results of (Q-LEACH) with existing protocols for WSNs. Moreover, MATLAB is used as a simulation tool.
\begin{figure}[!ht]
\centering
\subfigure[]{\includegraphics[height=5.5 cm,width=7 cm]{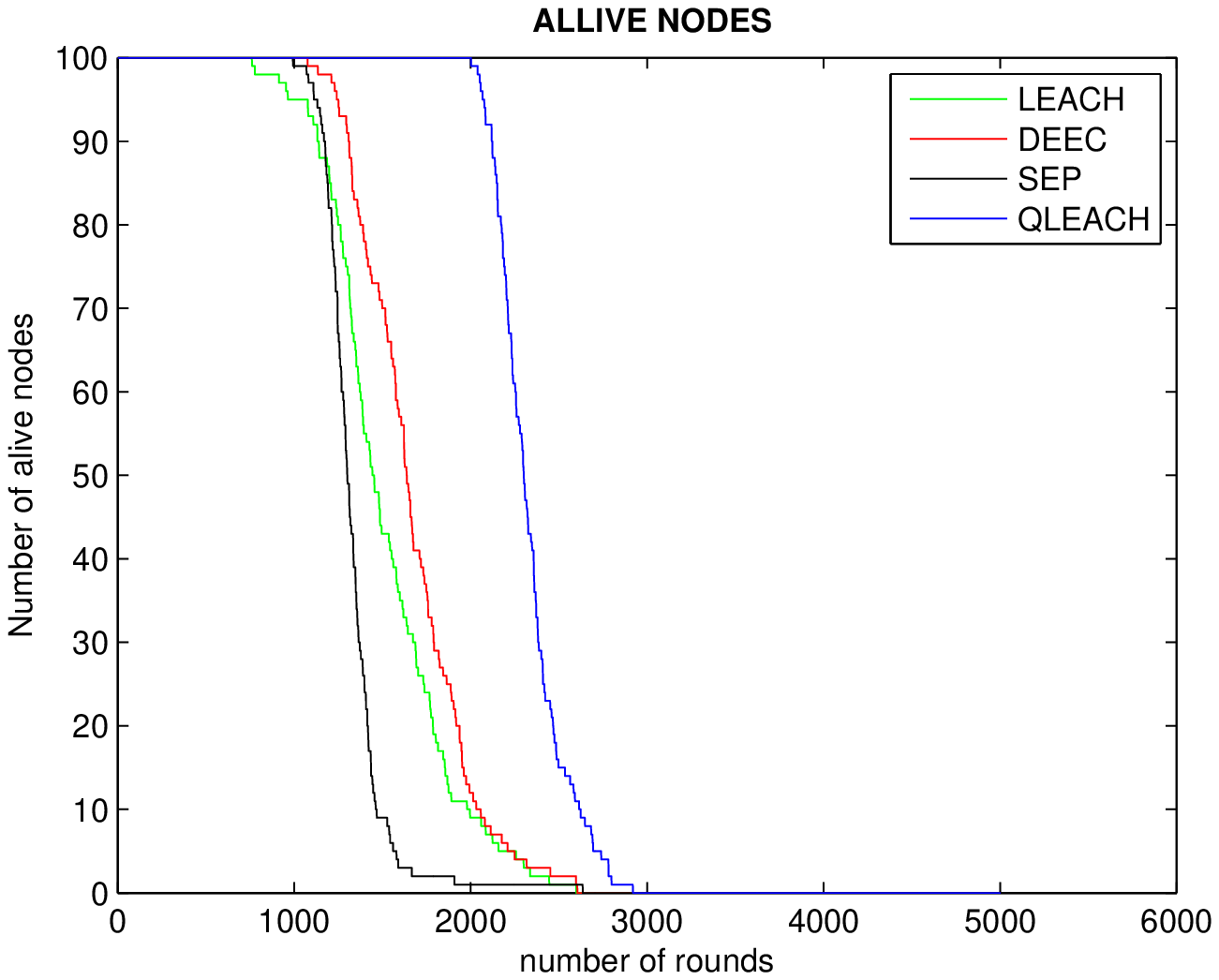}}
\subfigure[]{\includegraphics[height=5.5 cm,width=7 cm]{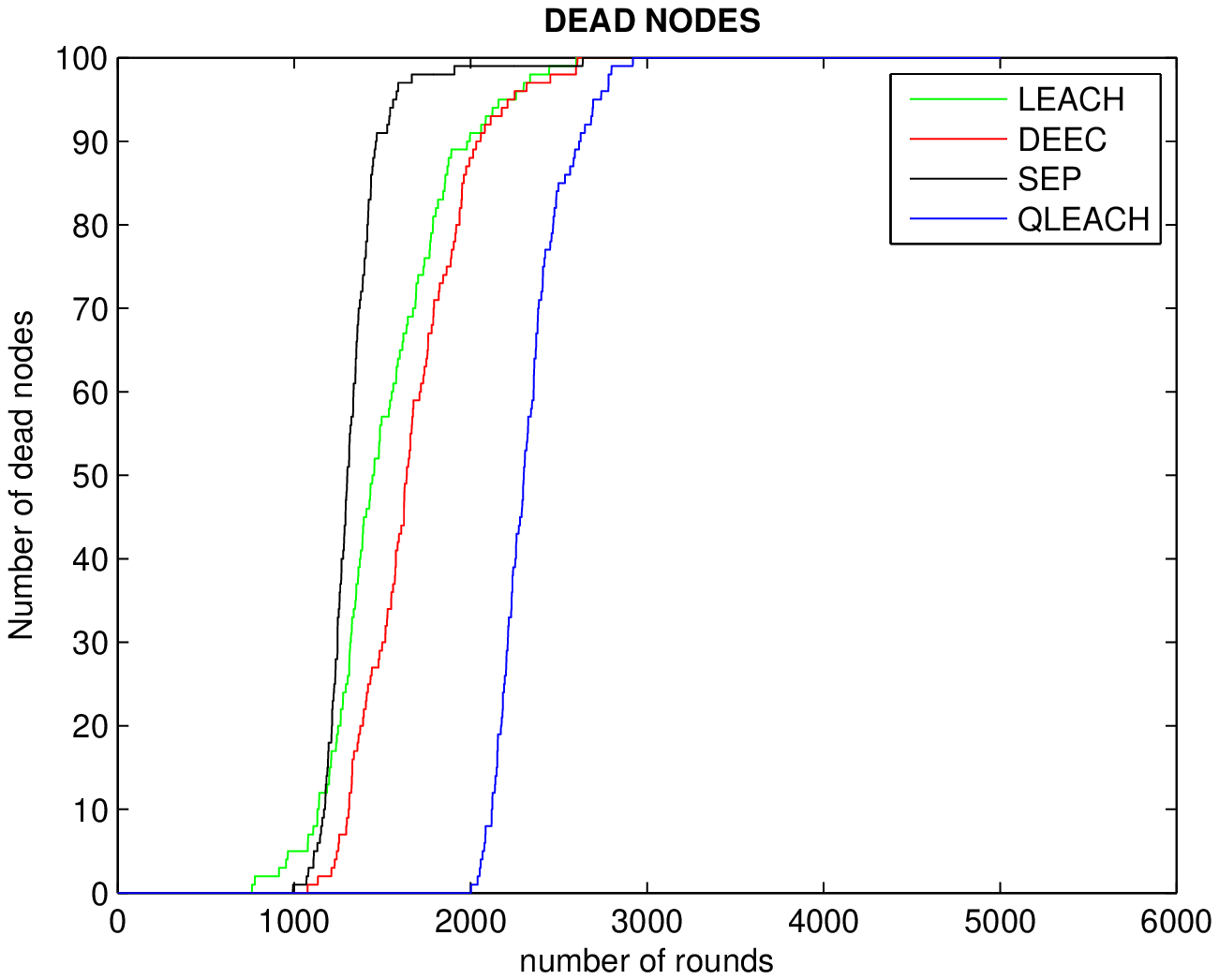}}
\subfigure[]{\includegraphics[height=5.5 cm,width=7 cm]{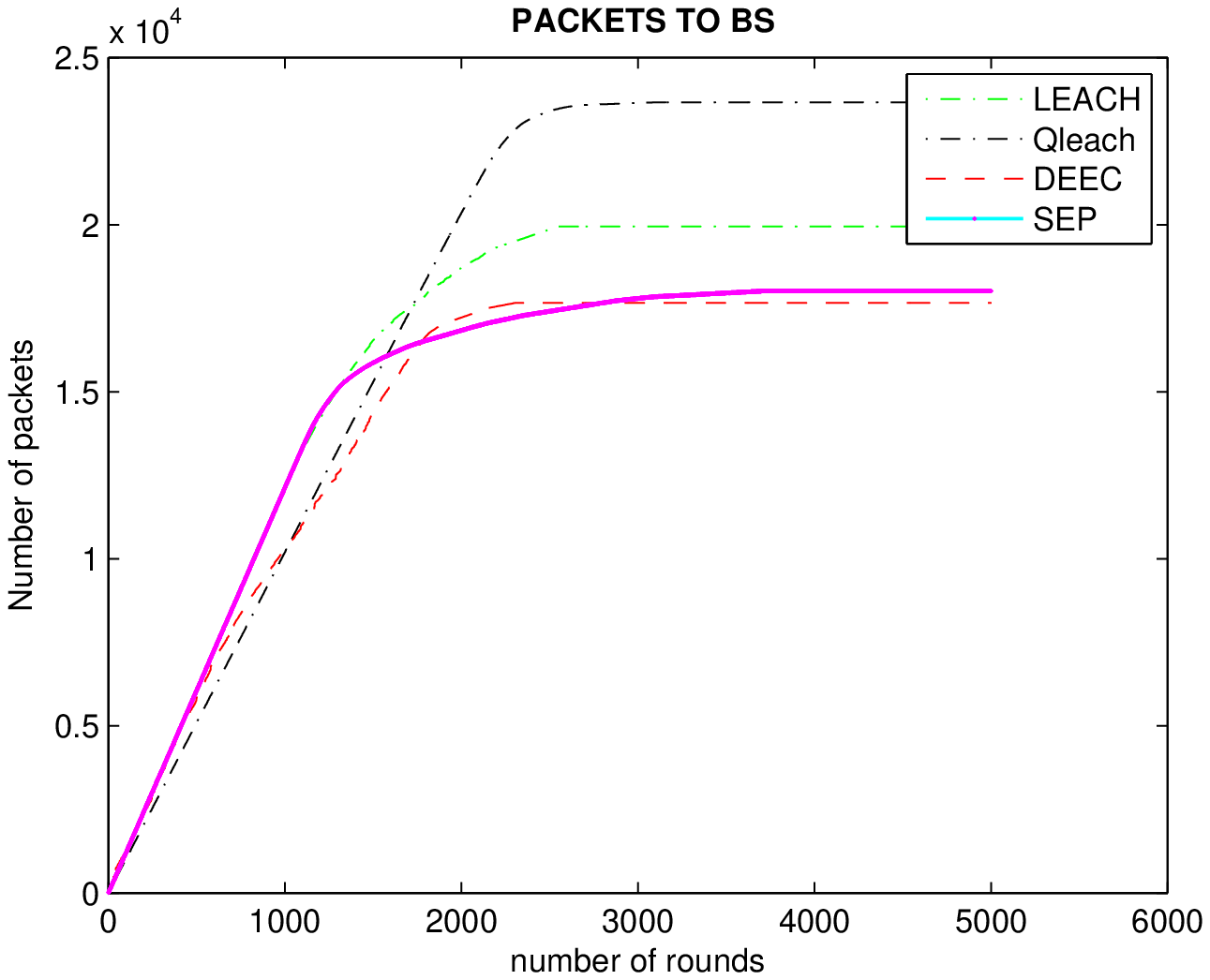}}
\subfigure[]{\includegraphics[height=5.5 cm,width=7 cm]{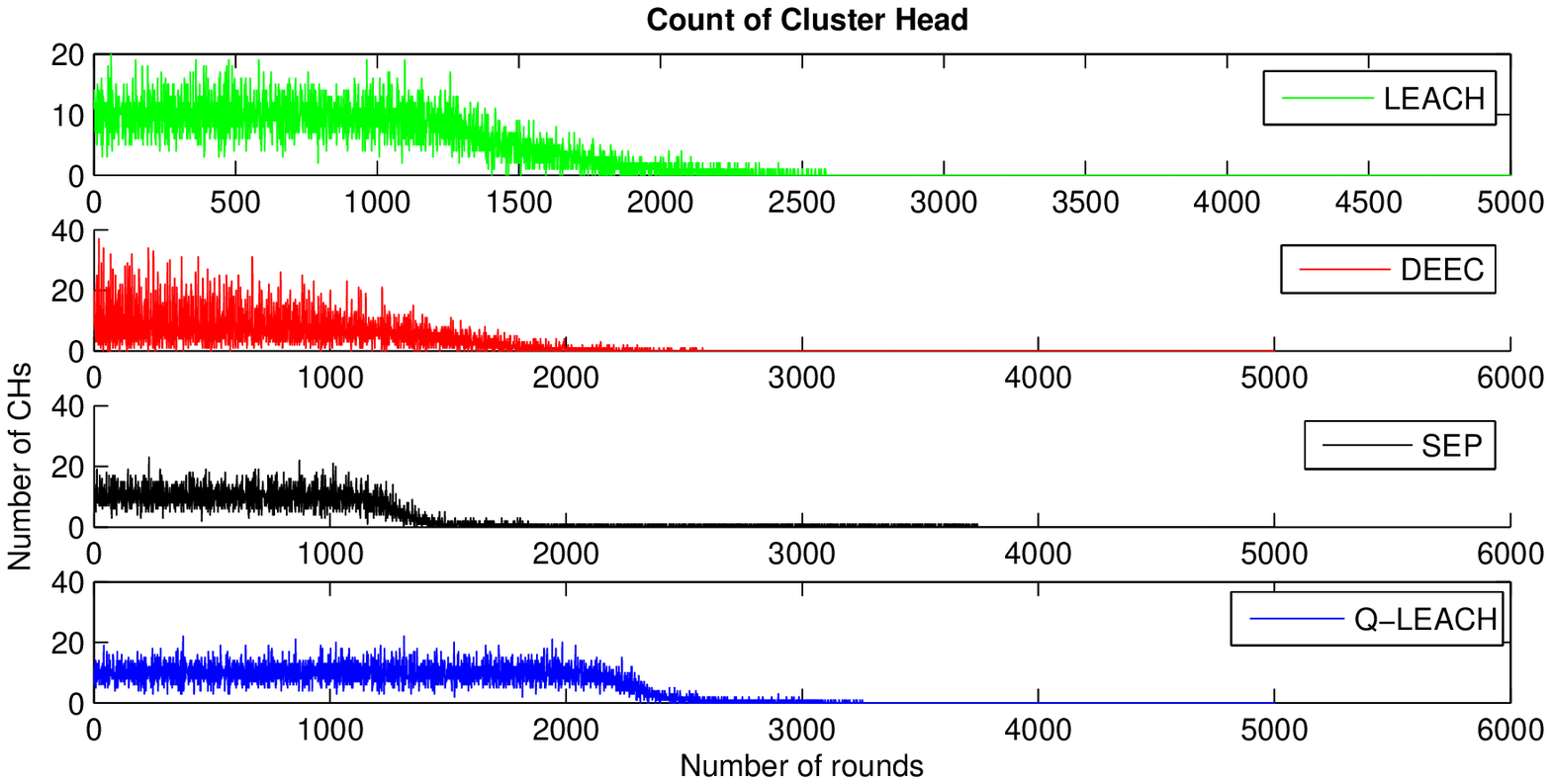}}
\subfigure[]{\includegraphics[height=5.5 cm,width=7 cm]{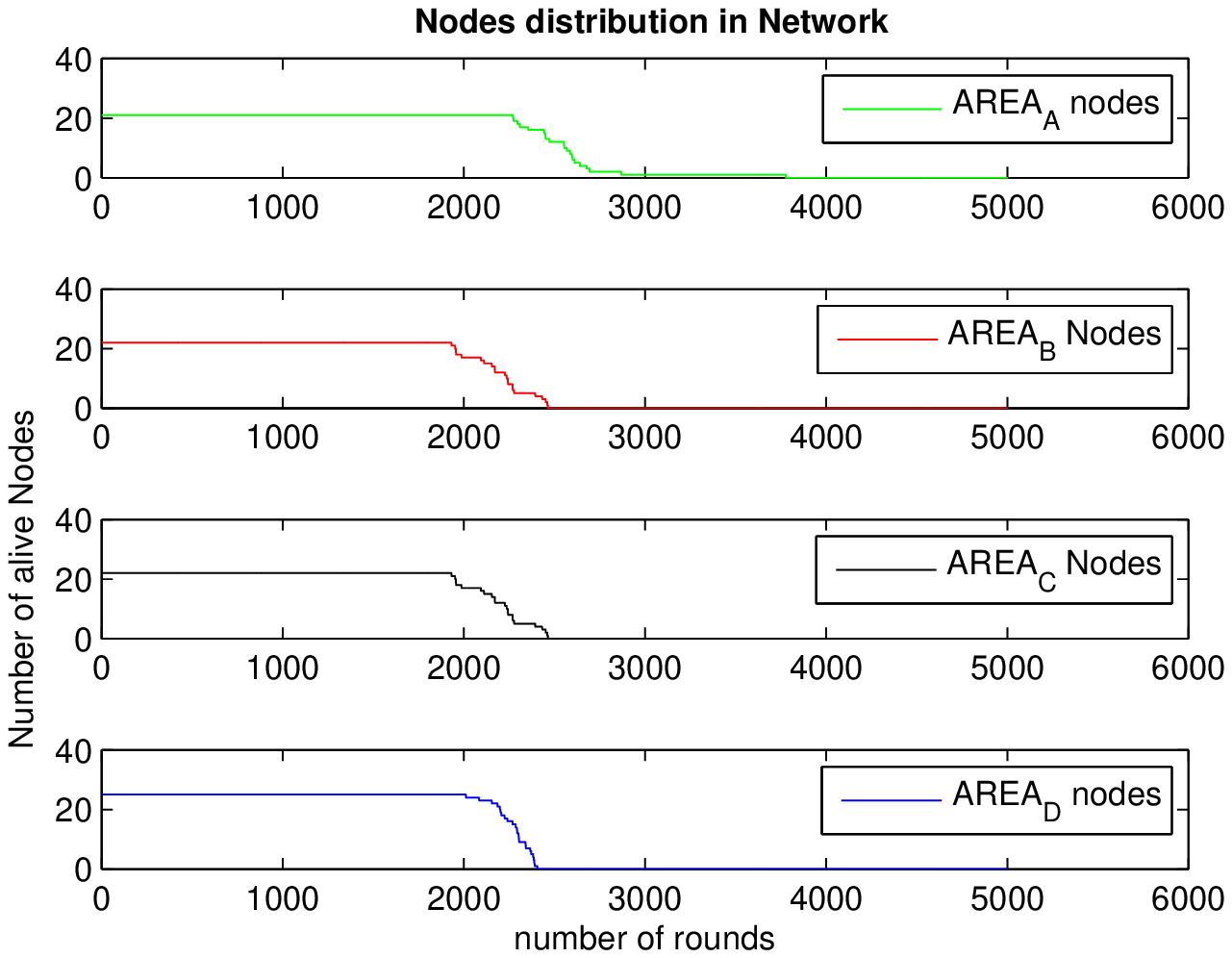}}
\caption{Performance Evaluation of LEACH, DEEC, SEP and Q-LEACH. }
\end{figure}

We deploy a random network of $100$ nodes with initial energy $=0.5j$ in filed with dimensions of $100m\times100m$. In simulated scenario BS is placed far away from the network field. We evaluated performance of our proposed strategy on the basis of certain parameter network stability period (S.P), network life time (N.L.T), and throughput (T.P).

In Fig 3.(a) it is shown that network life time is enhanced quiet significantly when compared with other clustering approaches i.e., when compared with LEACH, SEP, and DEEC, Q-LEACH performs better. In our case the network remains alive almost up to $2900$ rounds assuring network life-time to be more optimized. Moreover, it is also obvious that stability period is also improved i.e., first node dies around $2000$ rounds whereas, in schemes like LEACH, DEEC and SEP this value is much lower.

Fig 3.(b) represents unstable period of network when compared to other protocols it is clear that it also improved. Fig 3.(c) gives a picture of throughput and Fig 3.(d) represents CHs selected in every round for different clustering protocols. From results it is seen that selection of CHs follow some patterns and throughput increases quiet remarkably respectively. Fig 3.(e) shows the distribution of sensor nodes in different quadrants. From figure it is seen that nodes are distributed in a uniform order which makes clustering technique more effective and efficient. Moreover, study of nodes distribution in each sector aids in defining appropriate clusters.

\begin{table}[ht!]
  \centering
   \caption{Comparison of Network Parameters}
    \small
    \begin{tabular}{|p{1cm}|p{1.5cm}|p{1.5cm}|p{1.5cm}|p{1.5cm}|}
    \multicolumn{3}{c}{} \\
    \hline
    &\textbf{LEACH}&\textbf{DEEC}&\textbf{SEP}&\textbf{Q-LEACH}\\ \hline
    \textbf{S.P} & 700 & 1200 & 1000  & 2000 \\ \hline
    \textbf{N.L.T} & 1700  & 1900  &1800  & 2900  \\ \hline
    \textbf{T.P} &1735  & 17663  &13602   &23296 \\ \hline
    \end{tabular}
    \normalsize
\end{table}

 Table.1 defines comparison of network parameters. From table it is clear that Q-LEACH when compared with LEACH, DEEC, and SEP performs quiet well. Hence, enhances the network efficiency by a reasonable margin. In terms of S.P and N.L.T it shows improvement quiet significantly. Sufficient improvements in T.P are also observed when compared with LEACH, DEEC and SEP. Similarly N.L.T in proposed network model enhances quiet significantly. All these improvements are due to efficient clustering mechanism.

\section{Conclusion}

Many proposed clustering protocols for WSNs aimed at suitable energy utilization. Load balancing among sensor node is of key importance and it strictly depicts network life-time. In both homogenous and heterogenous networks, protocol design should be capable of best distribution. The main aim of this work is to enhance existing protocol such that more robust and optimized results can be achieved. Q-LEACH, significantly improved network parameters and seems to be an attractive choice for WSNs by extending and enhancing overall network quality parameters.


\begin{thebibliography}{00}

 \bibitem{heinzelman2000energy} W. Heinzelman, A. Chandrakasan, and H. Balakrishnan, “Energy-efficient communication protocol for wireless microsensor networks,” in System Sciences, 2000. Proceedings of the 33rd Annual Hawaii International Conference on, pp. 10–pp, IEEE, 2000.

 \bibitem{} S. Lindsey and C. Raghavendra, “Pegasis: Power-efficient gathering in sensor information systems,” in Aerospace conference
proceedings, 2002. IEEE, vol. 3, pp. 3–1125, IEEE, 2002.

 \bibitem{} O. Younis and S. Fahmy, “Heed: a hybrid, energy-efficient, distributed clustering approach for ad hoc sensor networks,” Mobile
Computing, IEEE Transactions on, vol. 3, no. 4, pp. 366–379, 2004.

 \bibitem{}G. Smaragdakis, I. Matta, and A. Bestavros, “Sep: A stable election protocol for clustered heterogeneous wireless sensor networks,”
tech. rep., Boston University Computer Science Department, 2004.

 \bibitem{} L. Qing, Q. Zhu, and M. Wang, “Design of a distributed energy-efficient clustering algorithm for heterogeneous wireless sensor
networks,” Computer communications, vol. 29, no. 12, pp. 2230–2237, 2006.

 \bibitem{} Y. Yu, R. Govindan, and D. Estrin, “Geographical and energy aware routing: A recursive data dissemination protocol for wireless
sensor networks,” tech. rep., Citeseer, 2001.

 \bibitem{} A. Elrahim, H. Elsayed, S. Ramly, and M. Magdy, “An energy aware wsn geographic routing protocol,” Universal Journal of
Computer Science and Engineering Technology, vol. 1, no. 2, pp. 105–111, 2010.

 \end{thebibliography}

\end{document}